\documentclass[11pt,twoside]{article}
\usepackage{asp2010}

\resetcounters

\begin{document}

\title{Characterizing the UV and X-ray Outflow in Mrk 509}
\author{
G. A. Kriss,$^{1,2}$ N. Arav,$^3$ J.S. Kaastra,$^{4,5}$ J. Ebrero,$^4$
          C. Pinto,$^4$ B. Borguet,$^3$ D. Edmonds,$^3$ E. Costantini,$^4$
          K. C. Steenbrugge,${6,7}$ R.G. Detmers,$^{4,5}$ E. Behar,$^8$
          S. Bianchi,$^9$ A. J. Blustin,$^{10}$ G. Branduardi-Raymont,$^{11}$
          M. Cappi,$^{12}$ M. Mehdipour,$^{11}$ P. Petrucci,$^{13}$ and G. Ponti$^{14}$
}
\affil{$^1$Space Telescope Science Institute,
        3700 San Martin Drive, Baltimore, MD, 21218, USA}
\affil{$^2$Department of Physics \& Astronomy, The Johns Hopkins University,
        Baltimore, MD, 21218, USA}
\affil{$^3$Department of Physics, Virginia Tech, Blacksburg, VA 24061, USA}
\affil{$^4$SRON Netherlands Institute for Space Research,
    Sorbonnelaan 2, 3584 CA Utrecht, The Netherlands}
\affil{$^5$Astronomical Institute, University of Utrecht, Postbus 80000, 3508
TA Utrecht, The Netherlands}
\affil{$^6$Instituto de Astronom\'ia, Universidad Cat\'olica del Norte, Avenida
Angamos 0610, Casilla 1280, Antofagasta, Chile}
\affil{$^7$Department of Physics, University of Oxford, Keble Road,
    Oxford OX1 3RH, UK}
\affil{$^8$Department of Physics, Technion, Haifa 32000, Israel}
\affil{$^9$Dipartimento di Fisica, Universita degli Studi Roma Tre, via della
Vasca Navale 84, I-00146 Roma, Italy}
\affil{$^{10}$Institute of Astronomy, University of Cambridge, Madingley Road,
    Cambridge CB3 0HA, UK}
\affil{$^{11}$Mullard Space Science Laboratory, University College London,
Holmbury St. Mary, Dorking, Surrey, RH5 6NT, UK}
\affil{$^{12}$INAF-IASF Bologna, Via Gobetti 101, I-40129 Bologna, Italy}
\affil{$^{13}$UJF-Grenoble 1 / CNRS-INSU, Institut de Plan\'etologie et d'Astrophysique
    de Grenoble (IPAG) UMR 5274, Grenoble, F-38041, France}
\affil{$^{14}$School of Physics and Astronomy, University of Southampton,
Highfield, Southampton SO17 1BJ, UK}

\begin{abstract}
We observed Mrk 509 during the fall of 2009 during a multiwavelength campaign
using XMM-Newton, Chandra, HST/COS, SWIFT, and Integral.
The 600-ks XMM/RGS spectrum finds two kinematic components and a discrete
distribution of ionized absorbers.  Our high S/N COS spectrum detects
additional complexity in the known UV absorption troughs from a variety of
sources in Mrk 509, including the outflow from the active nucleus,
the ISM and halo of the host galaxy, and infalling clouds or stripped gas
from a merger that are illuminated by the AGN. The UV absorption only partially
covers the emission from the AGN nucleus with covering fractions lower
than those previously seen with STIS, and are comparable to those seen with
FUSE. Given the larger apertures of COS and FUSE compared to STIS, we
favor scattered light from an extended region near the AGN as the explanation
for the partial covering. As observed in prior X-ray and UV spectra, the UV
absorption has velocities comparable to the X-ray absorption, but the bulk of
the ultraviolet absorption is in a lower ionization state with lower total
column density than the gas responsible for the X-ray absorption.
Variability compared to prior UV spectra lets us set limits on the location,
density, mass flux, and kinetic energy of the outflowing gas.
For component 1 at $-400~\rm km~s^{-1}$, the kinetic energy flux of both the
UV and the X-ray outflow is insufficient to have a significant impact on
further evolution of the host galaxy.
\end{abstract}

\section{Introduction}

Outflows from active galactic nuclei (AGN) may
have a significant impact on the evolution of their host galaxies
\citep{Silk98, Scannapieco04, Granato04, DiMatteo05, Hopkins08, Somerville08}.
Generally, to have an impact on the host galaxy's
evolution, models require the kinetic energy flux of the AGN outflow to be
$> $5\% of the bolometric luminosity of the AGN \citep{DiMatteo05}.
While some fraction of the outflows in low-luminosity AGN may not escape
their host galaxy, at least as measured in the local universe
\citep{Das05, Ruiz05, Das07},
the impact of the outflow on the host could be sufficient if the kinetic
energy flux is even ten times less, or $> 0.5$\% \citep{Hopkins10}.

Nearby AGN provide local analogs that can help us to understand the mechanics,
energetics, and chemical enrichment patterns that may play a significant
role in cosmic evolution at high redshift.
More than half of low-redshift AGN exhibit blue-shifted UV or X-ray
absorption features indicative of outflowing gas
\citep{Crenshaw03, Dunn07, Cappi09, Tombesi10}.
Understanding the geometry and the location of the outflow relative to the
active nucleus is a key to making an accurate assessment of the total mass
and the kinetic luminosity of the outflow.
Distance determinations are particularly difficult.
Using density-sensitive absorption lines to establish the gas density,
in combination with photoionization models that reproduce the observed
relative column densities can provide precise distance estimates.
These measures have ranged from tens of parsecs
in NGC 3783 \citep{Gabel05b} and NGC~4151
\citep{Kraemer06}, and up to kiloparsec scales in some quasars and AGN
\citep{Hamann01, Scott04a, Edmonds11}.

To improve upon these prior studies, we have conducted a multiwavelength
campaign of coordinated X-ray, UV, and optical observations of the nearby
luminous Seyfert 1 galaxy Mrk 509 (z=0.034397; \citealt{Huchra93}).
A complete overview of the campaign is given by \citet{Kaastra11a}.
Mrk 509 is an ideal object for study due to its high flux, moderate luminosity
that rivals that of QSOs \citep{Kopylov74}, and deep, well structured absorption
troughs.
Our observations using the Cosmic Origins Spectrograph (COS) \citep{Green11}
on HST obtained as part of this campaign
provide insight into long-term changes in the absorbing gas.

\section{UV Spectra of Mrk 509}

The COS spectra of Mrk 509 and their calibration are fully discussed by
\citet{Kriss11b}. The COS observations were obtained on
2009 December 10 and 11, and we compare these to the STIS echelle spectrum
obtained on 2001 April 13 by \citet{Kraemer03} and the FUSE spectrum obtained
by \citet{Kriss00}.
Figures 7 and 8 of \citet{Kriss11b} compare the normalized COS and STIS spectra.
\citet{Kriss11b} normalized the calibrated spectra by fitting an
emission model to the continuum and emission lines and dividing this model
into the calibrated spectra.
The same emission model was used for both STIS and COS with appropriately
fitted adjustments to the intensities of the emission line components and
the continuum.

In Fig. 7 of \citet{Kriss11b} one can see that components
1 and 1b in the COS Ly$\alpha$ spectrum are not as deep as in the STIS spectrum.
In contrast, component 1 in the N~{\sc v} absorption profile is noticeably
shallower in the STIS spectrum compared to the COS spectrum.
These changes are consistent with a change in ionization state of the
absorbing gas in response to the change in continuum flux.
The continuum flux seen in the COS observation is 80\% higher than for the
STIS observation. An increase in the ionization of the absorbing gas leads to
less neutral hydrogen and a lower column density for Ly$\alpha$, and increased
ionization and a higher column density for N~{\sc v}.

\section{Discussion}

The changes in the absorption seen between the STIS and the COS spectra enable
us to infer the density of the absorbing gas, and hence its distance, if we
assume that the variations are ionization changes in response to the changes
in ionizing flux.
Recombination times for hydrogen are too slow to permit us to set any
interesting limits on the density.
However, N~{\sc v} recombines nearly two orders of magnitude faster, and it provides
greater leverage in sensing the density of the absorbing medium.
Following \citet{KK95} and \citet{Nicastro99},
recombination and ionization time scales
depend not only on the density, but also on the relative populations of the
ionization states involved:
$\rm t_{rec} = (n_i / n_{i+1}) / (n_e \alpha_{rec})$.
Using the photoionization model for the Mrk 509 absorbers from
\citet{Kraemer03}, component 1 has an ionization parameter $\rm log~\xi = 0.67$
(where $\xi = \rm L_{ion} / (n r^2)$).
At this level, the ionization fraction for N~{\sc v} is $\sim 0.4$, and the
relative equality of populations among the neighboring ionization states of
nitrogen makes ionization and recombination timescales similar
\citep{Nicastro99}.
For $\alpha_{rec} = 8.96  \times 10^{-12}~cm^3~s^{-1}$ \citep{Nahar06}
at a temperature of 20,000 K, and a time between the STIS and COS observations
of $2.733 \times 10^{8}$ s, we get a lower limit on the density of
$\rm n_e > 160~cm^{-3}$. Since we are using the
photoionization models of \citet{Kraemer03}, we use an ionizing luminosity
from their SED in \citet{Yaqoob03},
$\rm L_{ion} = 7.0 \times 10^{44}~erg~s^{-1}$.
Together with the ionization parameter $\rm log~\xi = 0.67$,
this gives an upper limit on the distance of $\rm r < 250~pc$.

Just as limits on the variability timescale can give us lower limits on the
density and upper limits on the distance, the {\it lack} of variability in
Components 2--7 give us an {\it upper} limit on the density and a {\it lower}
limit on the distance. The paper by Edmonds et al. (2012) in this volume
shows in detail how the lack of variability in the other UV absorption
components results in a lower limit on their distance of $>$100--200 pc.

Overall, these results suggest that the UV absorption seen in our spectra of
Mrk 509 arise from a variety of locations in this active galaxy.
The high ionization and kinmatic correspondence of Components 1--3 suggest that
they are associated with the X-ray outflow, although the UV gas is lower in
ionization than the X-ray.
Similarly, Components 4a, 5-6 are associated with the lower-velocity portion of the X-ray outflow, but again, lower ionization than the X-ray gas.
The at-rest systemic velocity of Component 4 and its very low ionization
suggest that it is the ISM+Halo of the host galaxy.
Component 7, at a positive velocity of $200~\rm km~s^{-1}$, is infalling to
Mrk 509. We suggest that it might be similar to high-velocity clouds seen in the halo of our own galaxy. 
\citet{Thom08} find that HVC Complex C in the Milky Way has 
log n $\sim -2.5$, dimensions $3 \times 15$ kpc, distance of 10 kpc, and a
total mass of $8.2 \times 10^6~\rm M_\odot$.
For Component 7 in Mrk 509, for solar abundances, the total hydrogen column is
$\sim4 \times 10^{18}~\rm cm^{-2}$. If this is similar to Complex C, its size
is 1.3 kpc, and its distance from the center of Mrk 509 is 19 kpc.

The density and distance limits for Component 1 allow us to evaluate the mass
flux and kinetic luminosity:
{\center
$$\rm \dot M_{out} = 4\pi \Delta\Omega r N_H \mu m_p v_{out} = 3 \pi (r/250 pc)(N_H/1.0 \times 10^{19}~cm^{-2})(v/400~km~s^{-1})$$
$$\rm \dot M_{out} < 0.12 M_\odot~yr^{-1}, so$$
$$\rm L_k = 1/2 \dot M_{out} v_{out}^2 < 6.4 \times 10^{39} erg~s^{-1}$$
}

We measure $\rm L_{bol} = 6.4 \times 10^{45}~erg~s^{-1}$
(which requires $\rm \dot M_{acc} \sim 1.1 M_\odot~yr^{-1}$), so
$\rm L_k/L_{bol} < 1 \times 10^{-6}$. 

We note that even though the column density of the X-ray outflow is two orders
of magnitude higher \citep{Detmers11}, at the same velocity and distance it
would have a kinetic luminosity relative to bolometric of $< 10^{-4}$, still
more than an order of magnitude lower than needed to have a significant
impact on the host galaxy, even in the optimistic models of \citet{Hopkins10}.

What would constitute a significant outflow?
We need sufficient mass flux and velocity so that
$\rm L_k = 1/2 \dot M_{out} v_{out}^2 > 5\%~L_{rad} = 5\% \eta \dot M_{acc}c^2$,
where $\eta$ is the accretion efficiency.
If $\rm \dot M_{out} \sim \dot M_{acc}$, then
$\rm v_{out}^2 > 2 \times 5\% \eta c^2$.
Since $\eta \sim 0.1$, $\rm v_{out} > 0.1 c$.
Thus, even for a massive outflow in which the outflow rate is equal to the
accretion rate, outflow velocities would have to exceed $30,000~\rm km~s^{-1}$,
in order for the outflow to significantly influence the evolution of the host
galaxy. Even at a reduced efficiency of 0.5\% \citep{Hopkins10}, velocities
would have to be in excess of $10,000~\rm km~s^{-1}$. Such velocities are never
seen in local AGN, only in more distant broad absorption line quasars.
Allowing for an even more extreme case of the mass outflow exceeding the
accretion rate by an order of magnitude, outflow velocities would have to
be greater than 3,000 $~\rm km~s^{-1}$ for $\rm L_k/L_{bol} > 0.5\%$.
Velocities this high are very rare in local AGN.

In conclusion, we find that limits on the density and distance of the absorbers show that their kinetic luminosity is insufficient to cause significant feedback
affecting the evolution of the host galaxy.
Given that the crucial era for the effects of feedback, however, is during the
epoch of galaxy formation, the fact that we do find such significant influence
occuring at zero redshift is not a problem. Nothing rules out the possible
existence of stronger winds and outflows during the earlier lifetime of the
Mrk 509 host.

\acknowledgements 
This work was supported by NASA through grants for HST program number 12022
from the Space Telescope Science Institute, which is operated
by the Association of Universities for Research in Astronomy, Incorporated,
under NASA contract NAS5-26555,
and XMM-Newton grant NNX09AR01G from Goddard Space Flight Center.
SRON is supported financially by NWO,
the Netherlands Organization for Scientific Research.
KCS acknowledges the support of Comit\'e Mixto ESO - Gobierno de Chile.
SB acknowledges financial support from contract ASI-INAF n. I/088/06/0.
EB was supported by a grant from the Israel Science Foundation.
PoP acknowledges financial support from the GDR PCHE in France and
from the CNES French national space agency.
GP acknowledges support via an EU Marie Curie Intra-European Fellowship under
contract no. FP7-PEOPLE-2009-IEF-254279.

\bibliography{kriss_121611}

\begin{thebibliography}{}
\expandafter\ifx\csname natexlab\endcsname\relax\def\natexlab#1{#1}\fi
\expandafter\ifx\csname url\endcsname\relax
  \def\url#1{\texttt{#1}}\fi
\expandafter\ifx\csname urlprefix\endcsname\relax\def\urlprefix{URL }\fi
\providecommand{\eprint}[2][]{\url{#2}}

\bibitem[{{Cappi} et~al.(2009){Cappi}, {Tombesi}, {Bianchi}, {Dadina}
  et~al.}]{Cappi09}
{Cappi}, M., {Tombesi}, F., {Bianchi}, S., {Dadina}, M., et~al. 2009, \aap,
  504, 401

\bibitem[{{Crenshaw} et~al.(2003){Crenshaw}, {Kraemer}, \&
  {George}}]{Crenshaw03}
{Crenshaw}, D.~M., {Kraemer}, S.~B., \& {George}, I.~M. 2003, \araa, 41, 117

\bibitem[{{Das} et~al.(2005){Das}, {Crenshaw}, {Hutchings} et~al.}]{Das05}
{Das}, V., {Crenshaw}, D.~M., {Hutchings}, J.~B., et~al. 2005, \aj, 130, 945

\bibitem[{{Das} et~al.(2007){Das}, {Crenshaw}, \& {Kraemer}}]{Das07}
{Das}, V., {Crenshaw}, D.~M., \& {Kraemer}, S.~B. 2007, \apj, 656, 699

\bibitem[{{Detmers} et~al.(2011){Detmers}, {Kaastra}, {Steenbrugge}
  et~al.}]{Detmers11}
{Detmers}, R.~G., {Kaastra}, J.~S., {Steenbrugge}, K.~C., et~al. 2011, \aap,
  534, A38

\bibitem[{{Di Matteo} et~al.(2005){Di Matteo}, {Springel}, \&
  {Hernquist}}]{DiMatteo05}
{Di Matteo}, T., {Springel}, V., \& {Hernquist}, L. 2005, \nat, 433, 604

\bibitem[{{Dunn} et~al.(2007){Dunn}, {Crenshaw}, {Kraemer}, \&
  {Gabel}}]{Dunn07}
{Dunn}, J.~P., {Crenshaw}, D.~M., {Kraemer}, S.~B., \& {Gabel}, J.~R. 2007,
  \aj, 134, 1061

\bibitem[{{Edmonds} et~al.(2011){Edmonds}, {Borguet}, {Arav}, {Dunn}
  et~al.}]{Edmonds11}
{Edmonds}, D., {Borguet}, B., {Arav}, N., {Dunn}, J.~P., et~al. 2011, \apj,
  739, 7

\bibitem[{{Gabel} et~al.(2005){Gabel}, {Kraemer}, {Crenshaw} et~al.}]{Gabel05b}
{Gabel}, J.~R., {Kraemer}, S.~B., {Crenshaw}, D.~M., et~al. 2005, \apj, 631,
  741

\bibitem[{{Granato} et~al.(2004){Granato}, {De Zotti}, {Silva}, {Bressan}, \&
  {Danese}}]{Granato04}
{Granato}, G.~L., {De Zotti}, G., {Silva}, L., {Bressan}, A., \& {Danese}, L.
  2004, \apj, 600, 580

\bibitem[{{Green} et~al.(2011){Green}, {Froning}, {Osterman}, {Ebbets}
  et~al.}]{Green11}
{Green}, J.~C., {Froning}, C.~S., {Osterman}, S., {Ebbets}, D., et~al. 2011,
  ArXiv e-prints. \eprint{1110.0462}

\bibitem[{{Hamann} et~al.(2001){Hamann}, {Barlow}, {Chaffee} et~al.}]{Hamann01}
{Hamann}, F.~W., {Barlow}, T.~A., {Chaffee}, F.~C., et~al. 2001, \apj, 550, 142

\bibitem[{{Hopkins} et~al.(2008){Hopkins}, {Cox}, {Kere{\v s}}, \&
  {Hernquist}}]{Hopkins08}
{Hopkins}, P.~F., {Cox}, T.~J., {Kere{\v s}}, D., \& {Hernquist}, L. 2008,
  \apjs, 175, 390. \eprint{0706.1246}

\bibitem[{{Hopkins} \& {Elvis}(2010)}]{Hopkins10}
{Hopkins}, P.~F., \& {Elvis}, M. 2010, \mnras, 401, 7

\bibitem[{{Huchra} et~al.(1993){Huchra}, {Latham}, {da Costa}
  et~al.}]{Huchra93}
{Huchra}, J., {Latham}, D.~W., {da Costa}, L.~N., et~al. 1993, \aj, 105, 1637

\bibitem[{{Kaastra} et~al.(2011){Kaastra}, {Petrucci}, {Cappi}, {Arav}
  et~al.}]{Kaastra11a}
{Kaastra}, J.~S., {Petrucci}, P.-O., {Cappi}, M., {Arav}, N., et~al. 2011,
  \aap, 534, A36

\bibitem[{{Kopylov} et~al.(1974){Kopylov}, {Lipovetskii}, {Pronik}, \&
  {Chuvaev}}]{Kopylov74}
{Kopylov}, I.~M., {Lipovetskii}, V.~A., {Pronik}, V.~I., \& {Chuvaev}, K.~K.
  1974, Astrophysics, 10, 305

\bibitem[{{Kraemer} et~al.(2006){Kraemer}, {Crenshaw}, {Gabel}
  et~al.}]{Kraemer06}
{Kraemer}, S.~B., {Crenshaw}, D.~M., {Gabel}, J.~R., et~al. 2006, \apjs, 167,
  161

\bibitem[{{Kraemer} et~al.(2003){Kraemer}, {Crenshaw}, {Yaqoob}
  et~al.}]{Kraemer03}
{Kraemer}, S.~B., {Crenshaw}, D.~M., {Yaqoob}, T., et~al. 2003, \apj, 582, 125

\bibitem[{{Kriss} et~al.(2011){Kriss}, {Arav}, {Kaastra}, {Ebrero}
  et~al.}]{Kriss11b}
{Kriss}, G.~A., {Arav}, N., {Kaastra}, J.~S., {Ebrero}, J., et~al. 2011, \aap,
  534, A41

\bibitem[{{Kriss} et~al.(2000){Kriss}, {Green}, {Brotherton} et~al.}]{Kriss00}
{Kriss}, G.~A., {Green}, R.~F., {Brotherton}, M., et~al. 2000, \apjl, 538, L17

\bibitem[{{Krolik} \& {Kriss}(1995)}]{KK95}
{Krolik}, J.~H., \& {Kriss}, G.~A. 1995, \apj, 447, 512

\bibitem[{{Nahar}(2006)}]{Nahar06}
{Nahar}, S.~N. 2006, \apjs, 164, 280

\bibitem[{{Nicastro} et~al.(1999){Nicastro}, {Fiore}, \& {Matt}}]{Nicastro99}
{Nicastro}, F., {Fiore}, F., \& {Matt}, G. 1999, \apj, 517, 108

\bibitem[{{Ruiz} et~al.(2005){Ruiz}, {Crenshaw}, {Kraemer} et~al.}]{Ruiz05}
{Ruiz}, J.~R., {Crenshaw}, D.~M., {Kraemer}, S.~B., et~al. 2005, \aj, 129, 73

\bibitem[{{Scannapieco} \& {Oh}(2004)}]{Scannapieco04}
{Scannapieco}, E., \& {Oh}, S.~P. 2004, \apj, 608, 62

\bibitem[{{Scott} et~al.(2004){Scott}, {Kriss}, {Lee} et~al.}]{Scott04a}
{Scott}, J.~E., {Kriss}, G.~A., {Lee}, J.~C., et~al. 2004, \apjs, 152, 1

\bibitem[{{Silk} \& {Rees}(1998)}]{Silk98}
{Silk}, J., \& {Rees}, M.~J. 1998, \aap, 331, L1

\bibitem[{{Somerville} et~al.(2008){Somerville}, {Hopkins}, {Cox}
  et~al.}]{Somerville08}
{Somerville}, R.~S., {Hopkins}, P.~F., {Cox}, T.~J., et~al. 2008, \mnras, 391,
  481

\bibitem[{{Thom} et~al.(2008){Thom}, {Peek}, {Putman} et~al.}]{Thom08}
{Thom}, C., {Peek}, J.~E.~G., {Putman}, M.~E., et~al. 2008, \apj, 684, 364

\bibitem[{{Tombesi} et~al.(2010){Tombesi}, {Cappi}, {Reeves}
  et~al.}]{Tombesi10}
{Tombesi}, F., {Cappi}, M., {Reeves}, J.~N., et~al. 2010, \aap, 521, A57

\bibitem[{{Yaqoob} et~al.(2003){Yaqoob}, {McKernan}, {Kraemer}
  et~al.}]{Yaqoob03}
{Yaqoob}, T., {McKernan}, B., {Kraemer}, S.~B., et~al. 2003, \apj, 582, 105

\end{thebibliography}

\end{document}